\documentclass[pss,fleqn,embeddedheads]{w-art}
\usepackage{times}
\usepackage{w-thm}

\usepackage{fleqn,epsfig,color,graphicx,url,psfrag}
\usepackage{amsthm,amsfonts,amssymb,amsmath,esint}

\usepackage{geometry}
\usepackage{mathrsfs}
\usepackage{textcomp}
\usepackage{algorithm}
\usepackage{algorithmic}
\usepackage{bbm}

\theoremstyle{definition}\newtheorem{problem}{Problem}

\newcommand{\Field}[1]{{\bf{#1}}}
\newcommand{\Tensor}[1]{{\bf{#1}}}
\newcommand{\curl}{\nabla\times}

\newcommand{\curlk}{\MyColThreeVec{\partial_{x}}{\partial_{y}}{ik_{z}}\times}

\newcommand{\MyColThreeVec}[3]{\left(
  \begin{array}[c]{c}
    #1\\
    #2\\
    #3
  \end{array}\right)}

\def\myspace{\vspace{0.5cm}\noindent}
\newcommand{\hcurl}{\mathrm{H}\left(\curlw,\Omega\right)}
\newcommand{\curlw}{\mathbf{curl}\;}

\newcommand{\hcurlh}{V_{h}}

\usepackage[]{graphicx}
\begin{document}
\DOIsuffix{theDOIsuffix}
\Volume{XX}
\Issue{1}
\Copyrightissue{01}
\Month{01}
\Year{2010}
\pagespan{1}{}



\title[FEM Optical Simulation of Semiconductor Lasers]{Finite Element Simulation of the Optical Modes of Semiconductor Lasers}


\author[J. Pomplun]{Jan Pomplun\inst{1,2}\footnote{Corresponding
     author: e-mail: {\sf pomplun@zib.de}, Phone: +49\,30\,841\,85\,273}}
\author[]{Sven Burger\inst{1,2}}
\author[]{Frank Schmidt\inst{1,2}}
\author[]{Andrei Schliwa\inst{3}}
\author[]{Dieter Bimberg\inst{3}}
\author[]{Agnieszka Pietrzak\inst{4}}
\author[]{Hans Wenzel\inst{4}}
\author[]{G\"otz Erbert\inst{4}}
\address[\inst{1}]{
Zuse Institute Berlin,
Takustra{\ss}e 7,
14\,195 Berlin,
Germany
}
\address[\inst{2}]{
JCMwave GmbH,
Haarer Stra{\ss}e 14a,
85\,640 Putzbrunn,
Germany
}
\address[\inst{3}]{
Institut f\"ur Festk\"orperphysik,
TU Berlin, 
Hardenbergstra{\ss}e 36,
10\,623 Berlin,
Germany
}
\address[\inst{4}]{
Ferdinand-Braun-Institut f\"ur H\"ochstfrequenztechnik Berlin, 
Gustav-Kirchhoff-Stra{\ss}e 4,
12\,489 Berlin,
Germany
}

\begin{abstract}
In the present article we investigate optical near fields in semiconductor lasers. We perform finite element simulations for two different laser types, namely a super large optical waveguide (SLOW) laser, which is an edge emitter, and a vertical cavity surface emitting laser (VCSEL). We give the mathematical formulation of the different eigenvalue problems that arise for our examples and explain their numerical solution with the finite element method. Thereby, we also comment on the usage of transparent boundary conditions, which have to be applied to respect the exterior environment, e.g., the very large substrate and surrounding air.

For the SLOW laser we compare the computed near fields to experimental data for different design parameters of the device.
For the VCSEL example a comparison to simplified 1D mode calculations is carried out.
\end{abstract}
\maketitle                   

\section{Introduction}
The performance of novel nano-optical devices like SLOW lasers, or VCSELs depends very sensitively on the design of the system. Due to lack of practical experience and high fabrication costs of prototypes, numerical design and simulations are vital for the development of high performance devices. Numerical simulations can lead to optimized structures with, e.g., desired properties of the far or near field and also give insight into the physics of a device.

Since in modern devices and applications the wavelength of light is of the same order as the dimension of the simulated structures, Maxwell's equations have to be solved rigorously, in order to get accurate results for the electromagnetic field. Examples of such structures are semiconductor lasers, meta materials, photonic crystal devices, photolithographic masks and nano-resonators. A lot of different simulation techniques have been applied to and developed for nano-optical simulations, e.g., the finite element method (FEM), finite difference time domain simulations (FDTD), wavelet methods, finite integration technique (FIT), or rigorously coupled wave analysis (RCWA) \cite{HOF09}.

Here we present the finite element method for the solution of time harmonic Maxwell's equations in semiconductor laser devices. The finite element method offers the possibility for high order discretization schemes, which leads to very accurate results at low computational costs. In this work the FEM solver JCMsuite developed at the Zuse-Institute Berlin and JCMwave GmbH was used for simulations \cite{POM07,ZSC07,KAR09}.

\section{Mathematical formulation of eigenmode problems}

In our numerical analysis of semiconductor lasers we focus on an edge emitting laser and a VCSEL. Our goal is to compute optical near field distributions of the lasing modes within these devices. In practice these simulations can be used for the design of real world devices, e.g., to study the dependence of the light pattern emitted from the laser on geometrical parameters. As an illustration such an analysis will be carried out for an edge emitting laser and we will compare our results to experimental measurements in Section \ref{sec:slow}. The computed lasing eigenmodes are also a crucial ingredient for full electro-optical device simulation \cite{CHU09}. However, this will be part of future work.

Mathematically computation of modes for the edge emitting laser and the VCSEL example lead to different eigenvalue problems. For the SLOW laser a propagating mode problem has to be solved and a resonance problem for the VCSEL. We are not only interested in the near field distribution, but also want to determine leakage losses of the lasing modes. Therefore, we can not restrict the domain of interest, e.g., to a cavity of finite size. The eigenvalue problems are per se stated on unbounded domains, which take into account the infinite surrounding exterior, i.e., the substrate and surrounding air. Since in a numerical simulation the size of the computational domain has to be restricted, we have to apply appropriate transparent/radiating boundary conditions. This is done by formulating the eigenvalue problem as a coupled interior-exterior problem.

We are interested in stationary solutions for the electric field $\Field E$. Therefore, we make the following harmonic ansatz for the time dependence:
\begin{align}
  \Field E(x,y,z,t)&=e^{-i\omega t}\Field E(x,y,z),
\end{align}
where $\omega$ denotes the frequency. Using this ansatz in Maxwell's equations, the following second order ``curl-curl equation'' for the electric field can be derived:
\begin{align}
  \label{eq:mwE}
\curl\Tensor{\mu}^{-1}\curl \Field{E}-\omega^{2}\Tensor{\varepsilon}\Field{E}&=0,
\end{align}
where we assumed no exterior sources. The quantities $\Tensor{\mu}$ and $\Tensor{\varepsilon}$ denote the permeability and permittivity tensor respectively.

This formulation is used in the following for derivation of the eigenvalue problems.

\subsection{Propagating mode problem}
\begin{figure}[t]
\psfrag{x}{$x$}
\psfrag{y}{$y$}
\psfrag{z}{$z$}
\psfrag{gamma}{$\Gamma$}
\psfrag{omega}{$\Omega_{\mathrm{in}}$}
\hspace{2cm}(a)\hspace{8cm}(b)\hfill\\\noindent
\includegraphics[height=3.6cm]{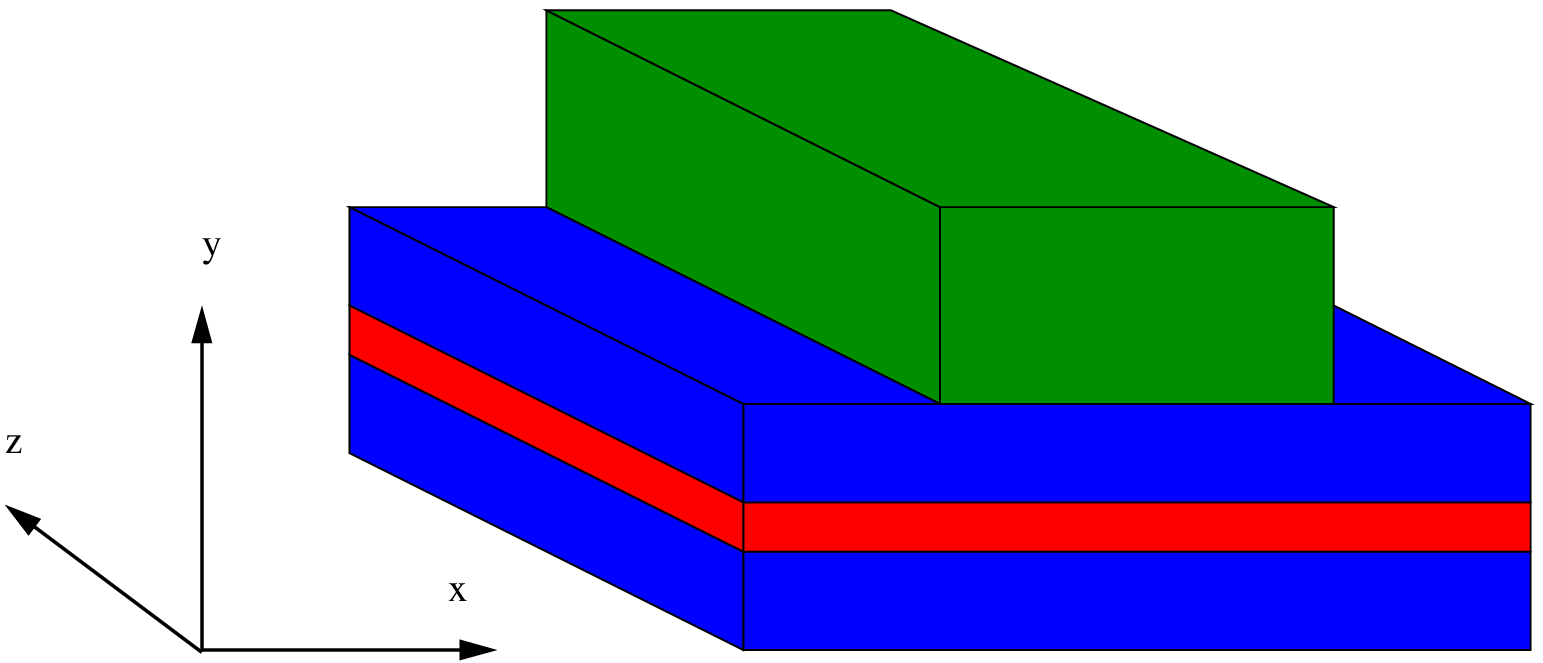}\hfill
\includegraphics[height=3.2cm]{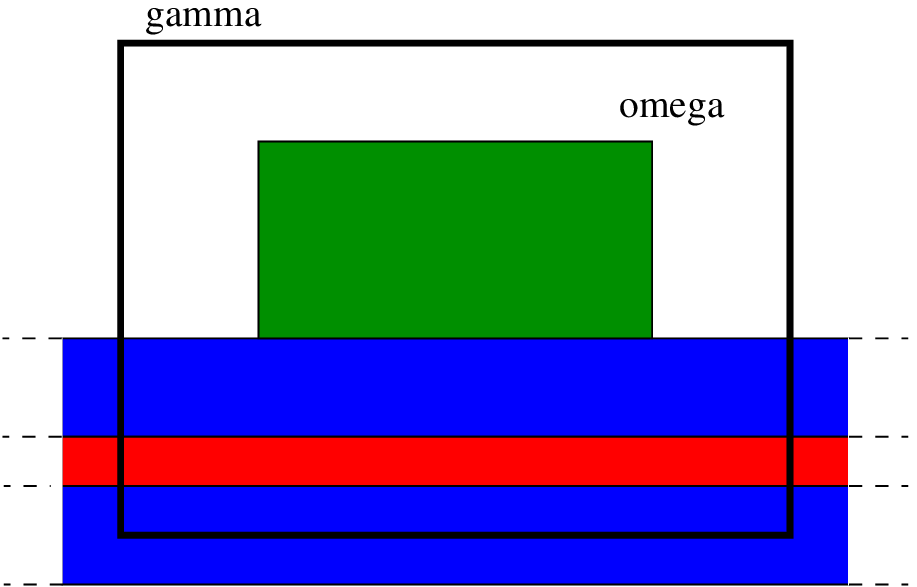}
\caption{\label{fig:pm}(a) Geometry of a propagating mode problem with invariance in $z$-direction; (b) corresponding 2D cross section. The geometry is divided into an interior domain $\Omega_{\mathrm{in}}$ with boundary $\Gamma$. The structured exterior domain extends to infinity.}
\end{figure}
Propagating mode problems arise, when waveguide structures as depicted in Fig. \ref{fig:pm}(a) are considered. The geometry has an invariance in one spatial dimension (here we choose the $z$-direction). Then the following ansatz for the electric field can be made:
\begin{align}
\label{eq:ansatzPM}
  E(x,y,z)&=e^{i k_{z} z}E(x,y),
\end{align}
where $k_{z}$ is the propagation constant. Hence, we assume a harmonic dependence of the electric field in the invariant direction. Note that with ansatz \eqref{eq:ansatzPM}, the differential operator with respect to $z$ can be replaced as follows:
\begin{align}
  \partial_{z}\equiv ik_{z}.
\end{align}
With ansatz \eqref{eq:ansatzPM}, furthermore, we only have to consider the cross section of the waveguide in the $xy$-plane. Figure \ref{fig:pm}(b) depicts this infinite domain of interest. The domain is divided into a finite interior $\Omega_{\mathrm{in}}\subset\real^{2}$ and an infinite exterior $\real^{2}\setminus\Omega_{\mathrm{in}}$. The boundary of $\Omega_{\mathrm{in}}$ is denoted by $\Gamma$. The propagating mode problem can now be given as follows:
\begin{problem}[Propagating mode problem]
\label{pr:pmStrong}
  For given frequency $\omega$ find pairs $\left(\left[\Field E_{\mathrm{in}},\Field E_{\mathrm{ex}}\right],k_{z}\right)$ such that:
  \begin{itemize}
  \item[(i)] The electric field $\Field E_{\mathrm{in}}$ fulfills Maxwell's equations in the interior domain:
\begin{align}
  \label{eq:mwInt}
\curlk\Tensor{\mu}^{-1}\curlk \Field E_{\mathrm{in}}(x,y)-\omega^{2}\Tensor{\varepsilon}\Field E_{\mathrm{in}}(x,y)&=0\;\mbox{in $\Omega_{\mathrm{in}}$}.
\end{align}
  \item[(ii)] The electric field $\Field E_{\mathrm{ex}}$ fulfills Maxwell's equations in the exterior domain:
\begin{align}
  \label{eq:mwExt}
\curlk\Tensor{\mu}^{-1}\curlk \Field E_{\mathrm{ex}}(x,y)-\omega^{2}\Tensor{\varepsilon}\Field E_{\mathrm{ex}}(x,y)&=0\;\mbox{in $\real^{2}\setminus\Omega_{\mathrm{in}}$}.
\end{align}
  \item[(iii)] Boundary condition at $\Gamma$: the tangential component of the electric field is continuous:
    \begin{align*}
      \Field{n}\times\left.\left(\Field E_{\mathrm{in}}-\Field E_{\mathrm{ex}}\right)\right|_{\Gamma}=0,
    \end{align*}
where \Field{n} denotes the outward pointing normal vector on $\Gamma$.
\item[(iv)]
Radiating boundary condition: $\Field E_{\mathrm{ex}}$ is strictly outward radiating.\\\noindent
For homogeneous exteriors this can be stated, e.g., by the Silver-M\"uller radiation condition:
\begin{align*}
&&\lim\limits_{r\rightarrow\infty}r\left( \curl \Field E_{\mathrm{sc}}(\Field{r})\times\Field{r}_{0}-i\frac{\omega\sqrt{\epsilon_{\mathrm{ext}}\mu_{\mathrm{ext}}}}{c}\curl \Field E_{\mathrm{sc}}(\Field{r})  \right)=0\\
&&\mbox{uniformly continuous in each direction $\Field{r}_{0}$,}
\end{align*}
where $\Field{r}$ is the coordinate vector in $\real^{2}$, $r$ its norm, $\Field{r}_{0}=\frac{\Field{r}}{r}$, and $\epsilon_{\mathrm{ext}}$ and $\mu_{\mathrm{ext}}$ the relative permittivity and permeability in the exterior.\\\noindent
A more general characterization of outward radiating fields for structured exterior domains is the pole condition \cite{Schmidt02H,Hohage01a}.
  \end{itemize}
\end{problem}\myspace
Problem \ref{pr:pmStrong} yields a quadratic eigenvalue problem for the electric field and the propagation constant $k_{z}$. The division of the electric field into an interior and exterior field has to be done, in order to guarantee that the computed modes are outward radiating. Physically this means that electromagnetic field energy of a mode within the waveguide only propagates outwards. Since the propagating mode problem is stated on an unbounded domain, the eigenvalue $k_{z}$ becomes complex. Hence, the propagating mode looses energy while propagating in $z$-direction and is damped according to:
\begin{align*}
  \Field E(x,y,z)&=e^{-\Im\{k_{z}\} z}\left(e^{i \Re\{k_{z}\} z}\Field E(x,y)\right),
\end{align*}
where $\Re\{k_{z}\}$ and $\Im\{k_{z}\}$ denotes the real and imaginary part of the propagation constant $k_{z}$. In numerical simulations we will compute the effective refractive index:
\begin{align}
  n_{\mathrm{eff}}=\frac{k_{z}}{k_{0}},
\end{align}
instead of the propagation constant $k_{z}$. Thereby $k_{0}=\omega/c$ is the vacuum wave number.
\subsection{Resonance problem}
The setup of a resonance problem differs from the propagating mode problem. Here, the electric field and its resonance frequency have to be determined for a given geometry, which can be 2D or 3D. Again the domain of interest is unbounded, and it is divided into an interior $\Omega_{\mathrm{in}}\subset\real^{3}$ and an exterior $\Omega_{\mathrm{in}}\setminus\real^{3}$, analogously to Fig. \ref{fig:pm}. The coupled interior-exterior resonance problem is given as follows:
\begin{problem}[Resonance mode problem]
\label{pr:resStrong}
  Find pairs $\left(\left[\Field E_{\mathrm{in}},\Field E_{\mathrm{ex}}\right],\omega\right)$ such that:
  \begin{itemize}
  \item[(i)] The electric field $\Field E_{\mathrm{in}}$ fulfills Maxwell's equations in the interior domain:
\begin{align}
  \label{eq:mwIntRes}
\curl\Tensor{\mu}^{-1}\curl \Field E_{\mathrm{in}}(x,y,z)-\omega^{2}\Tensor{\varepsilon}\Field E_{\mathrm{in}}(x,y,z)&=0\;\mbox{in $\Omega_{\mathrm{in}}$}.
\end{align}
  \item[(ii)] The electric field $\Field E_{\mathrm{ex}}$ fulfills Maxwell's equations in the exterior domain:
\begin{align}
  \label{eq:mwExtRes}
\curl\Tensor{\mu}^{-1}\curl \Field E_{\mathrm{ex}}(x,y,z)-\omega^{2}\Tensor{\varepsilon}\Field E_{\mathrm{ex}}(x,y,z)&=0\;\mbox{in $\real^{3}\setminus\Omega_{\mathrm{in}}$}.
\end{align}
  \item[(iii)] Boundary condition at $\Gamma$: the tangential component of the electric field is continuous:
    \begin{align*}
      \Field{n}\times\left.\left(\Field E_{\mathrm{in}}-\Field E_{\mathrm{ex}}\right)\right|_{\Gamma}=0.
    \end{align*}
\item[(iv)]
Radiating boundary condition: $\Field E_{\mathrm{ex}}$ is strictly outward radiating.
  \end{itemize}
\end{problem}\myspace
Problem \ref{pr:resStrong} is an eigenvalue problem for the frequency $\omega^{2}$ and the electric field. Again the division of the electric field is necessary to obtain outward radiating resonances. Since the resonance problem is stated on an infinite domain, there are no real-valued eigenvalues. The eigenfrequencies become complex, and their imaginary parts correspond to the inverse lifetimes and radiation losses of the corresponding eigenmodes:
\begin{align*}
  \Field E(x,y,z,t)&=e^{-\tau t}\left( e^{i\Re\{\omega\} t}\Field E(x,y,z)\right)\,,\;\mbox{with}\;\tau=-\frac{1}{\Im\{\omega\}}.
\end{align*}

\section{Transparent boundary conditions}
The eigenvalue problems introduced in the previous section are stated on unbounded domains. This makes straightforward discretization impossible, since we can only consider finite domains in a numerical simulation. Here we use the perfectly matched layer (PML) method \cite{BerPML} to reduce the domain of interest to finite size. Hence, the PML can be seen as a transparent boundary condition. 

The key idea of the PML method is a complex continuation of the electric field. Let us consider a 1D setup, where the positive real line is the infinite exterior. An outward radiating field, which travels from the origin $x=0$ to the right, is given by a plane wave:
\begin{align*}
  E(x)&=e^{ikx}\,,\;\mbox{with}\;k>0.
\end{align*}
Now we introduce a new coordinate system as follows:
\begin{align*}
   x=(1+i\sigma)\tilde x\,,\;\mbox{with}\;0<\sigma\in\real.
\end{align*}
The plane wave in the new coordinate system is given by:
\begin{align*}
  \tilde E(\tilde x)&=e^{-\sigma k\tilde x}e^{ik\tilde x},
\end{align*}
hence, it is decaying exponentially with increasing distance from the origin. Therefore, we can set $\tilde E(\tilde x)=0$ for a sufficiently large $\tilde x$. This means we can truncate the exterior domain at a finite value. 

In 2D and 3D the implementation of this idea is more technical \cite{Zschiedrich03}. First a coordinate system in the exterior has to be introduced, which includes a generalized distance variable $\xi$. Then a complex continuation along this distance variable is carried out, and the exterior domain is truncated after sufficient decay of the complex continuation of the exterior field. At this artificial boundary the electric field or its derivative are set to zero, corresponding to zero Dirichlet or Neumann conditions. This guarantees that the exterior field is strictly outward radiating.

\section{Finite element discretization}
\begin{figure}[t]
\psfrag{xl}{}
\psfrag{yl}{}
\includegraphics[height=4cm]{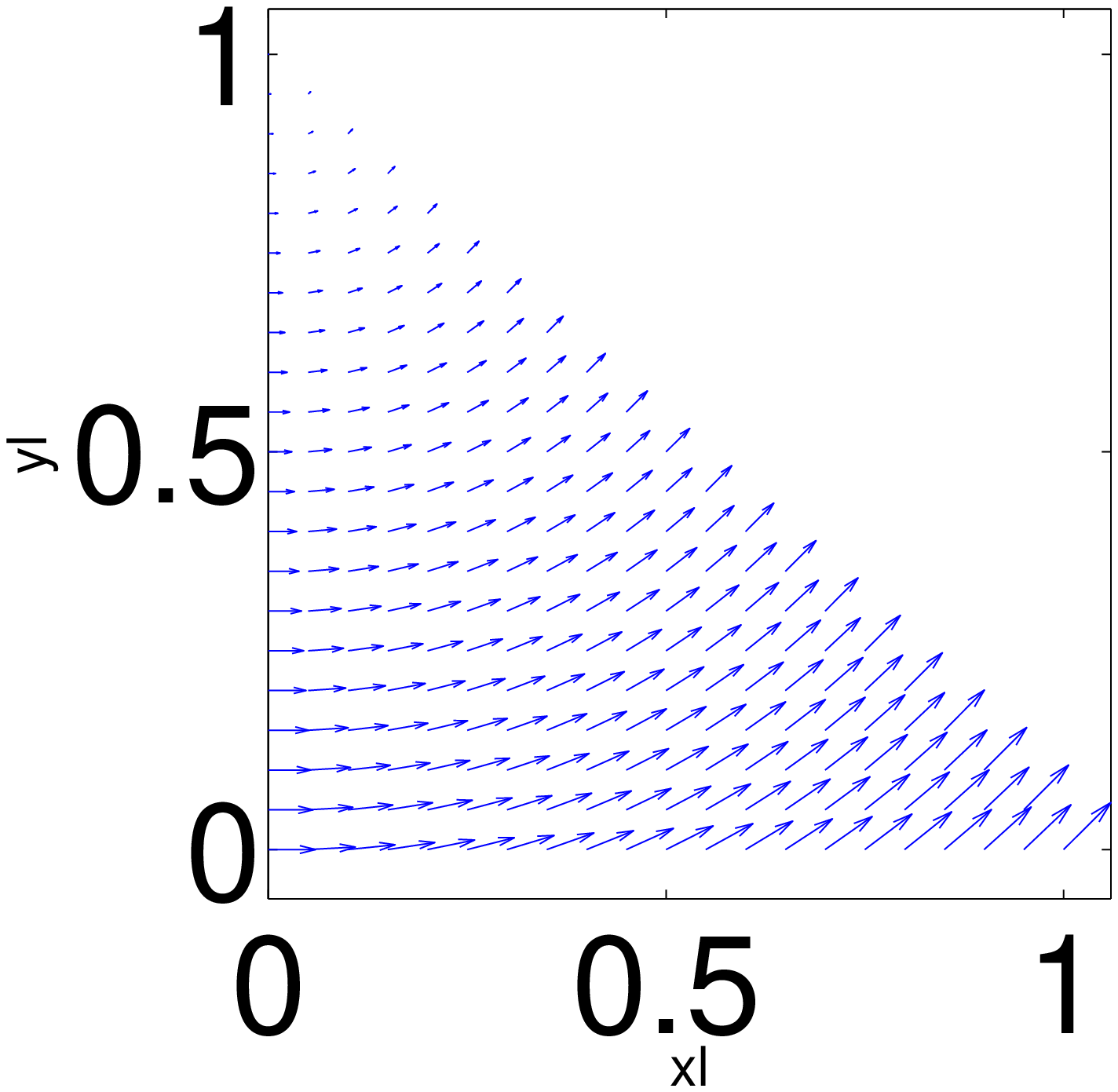}\hfill
\includegraphics[height=4cm]{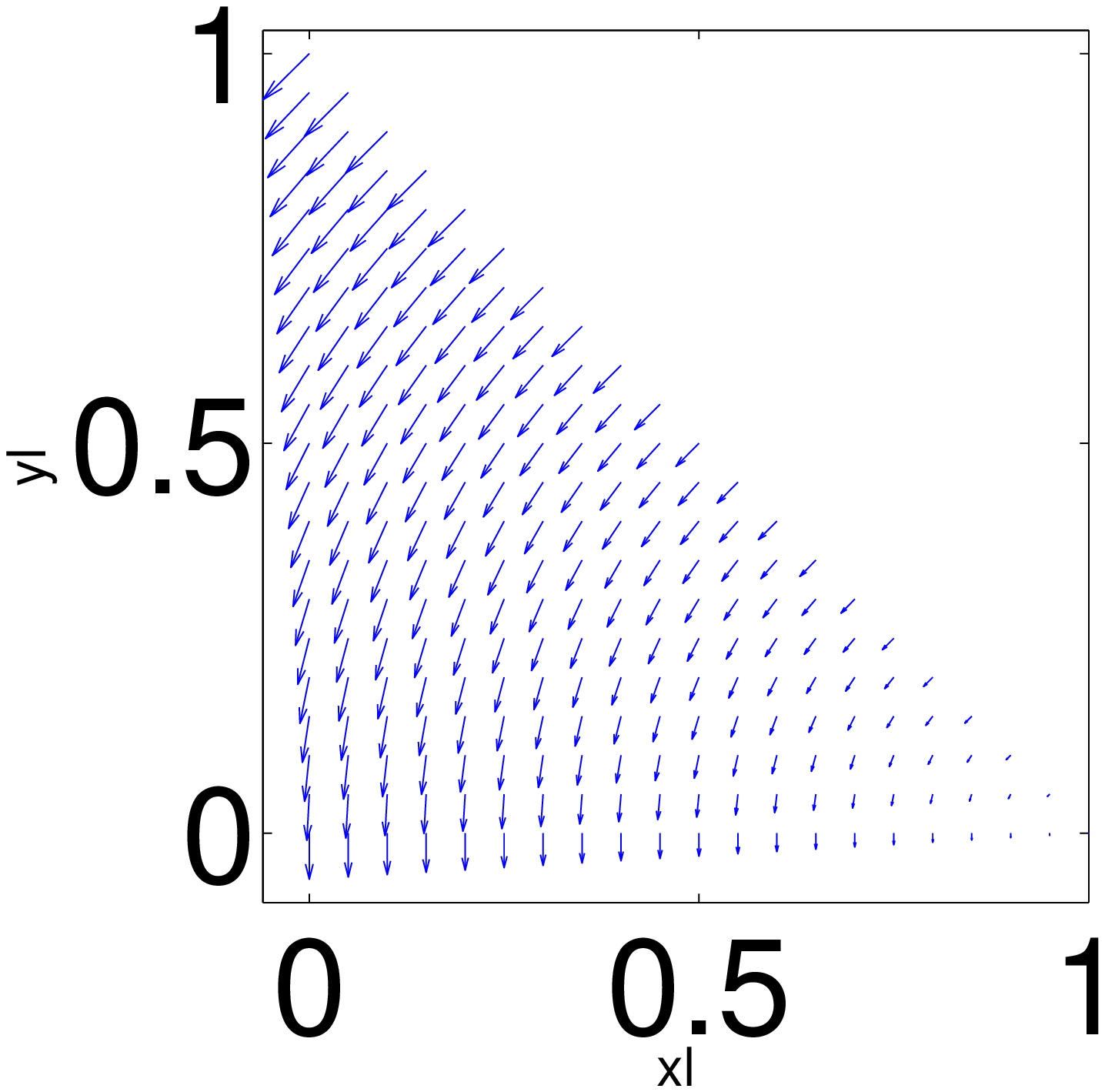}\hfill
\includegraphics[height=4cm]{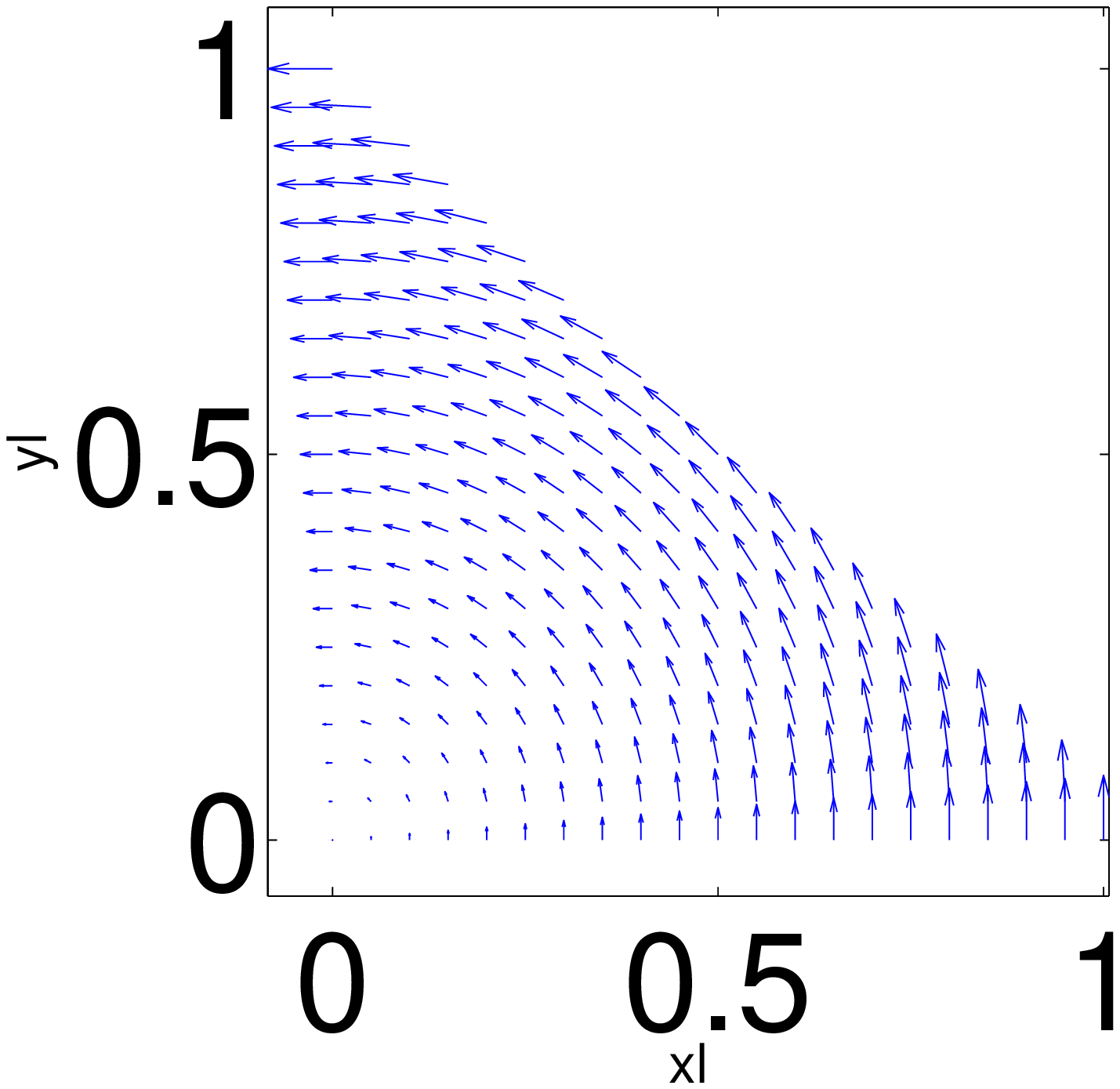}
\caption{\label{fig:fem}First order vectorial ansatz functions defined on triangle of finite element discretization.}
\end{figure}
For finite element discretization the curl-curl equation for the electric field \eqref{eq:mwE} has to be stated in so-called weak form:\\\noindent
Find $\Field E\in \hcurl$, such that:
\begin{align*}
  a(\Field E,v)&=f(v)\,,\;\forall v\in \hcurl.
\end{align*}
This is a variational formulation of Maxwell's equations, where the electric field $\Field E$ is determined from a proper infinite dimensional function space $\hcurl$ \cite{MON03}, and the bilinear form $a(\cdot,\cdot)$ basically represents the curl-curl equation. This offers a very elegant discretization scheme. The finite element version of Maxwell's equations simply reads:\\\noindent
Find $\Field E\in \hcurlh$, such that:
\begin{align*}
  a(\Field E,v)&=f(v)\,,\;\forall v\in \hcurlh,
\end{align*}
where the variational formulation is stated on a finite dimensional function space $\hcurlh\subset\hcurl$, with $\dim \hcurlh=N<\infty$. The finite element space $\hcurlh$ is constructed as follows. First the domain of interest is subdivided into a number of small patches, i.e., it is triangulated. Then on each patch a basis of polynomial ansatz functions of maximum polynomial degree $p$ is chosen, see Fig. \ref{fig:fem}. The union of all ansatz spaces on all patches gives the finite element space. 

The quality of a finite element solution can be increased by refining the patches, i.e., the mesh. This is referred to as $h$-refinement, where $h$ is the measure of the dimension of the patches. Another alternative is increasing the polynomial degree of the ansatz functions on each patch, which is referred to as $p$-refinement. For smooth solutions in general $p$-refinement (high order) leads to much faster convergence. We will analyze this in numerical examples.

\section{Finite element simulation of semiconductor lasers}
In this section we use the finite element method for computation of lasing modes of two types of semiconductor lasers. The first example is a 3D cylinder symmetrical VCSEL. We will compare our results to 1D simulations, which are often used in the design process due to their computational simplicity. 

The second example is a SLOW laser, which is produced at the Ferdinand Braun Institute (FBH). For this device experimental data of the near field distributions of the lasing modes is available, which can be compared to numerical results. The influence of design parameters is studied, and the finite element simulations will provide an explanation of the experimentally observed behaviour. Further we will analyze the convergence of the finite element simulation for this example.

For the waveguide problem \ref{pr:pmStrong} the frequency $\omega$ of the mode is fixed. Hence the dispersion of materials is not important. For the resonance mode problem the frequency $\omega$ is the unknown eigenvalue. Since the material parameters $\varepsilon$ depend on $\omega$ this leads to a non-linear eigenvalue problem in general, which can be solved iteratively. Usually the DBR mirrors of a VCSEL are designed for a specific frequency. This frequency is used as an initial guess to fix the material parameters. After the eigenvalue $\omega$ is computed, the material parameters are determined for the new frequency value. This procedure is continued until the eigenvalue $\omega$ is converged. 

\subsection{Vertical cavity surface emitting laser}
\begin{figure}[t]
\centering
\includegraphics[width=12cm]{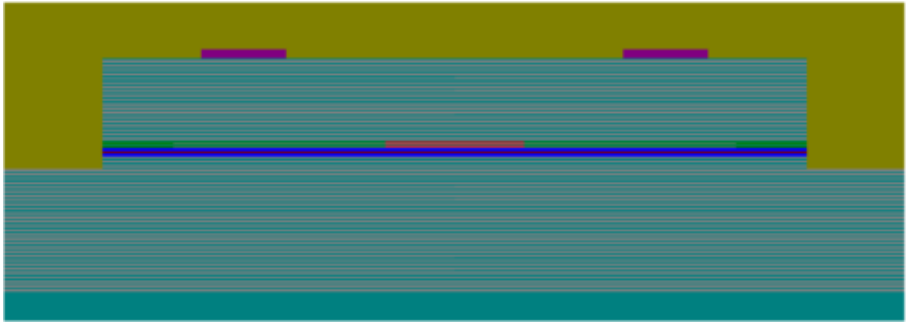}\hfill
\caption{\label{fig:vcselGeo}Cross section of cylinder symmetrical VCSEL.}
\end{figure}
\begin{figure}[]
\centering
\includegraphics[clip,width=12cm]{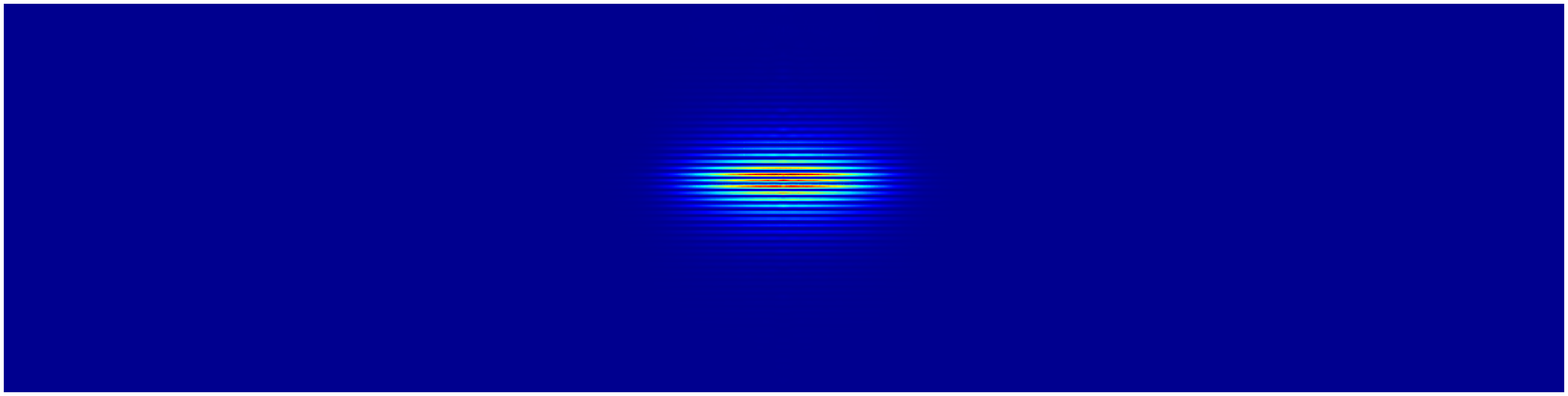}\vspace{0.5cm}
\includegraphics[width=12cm]{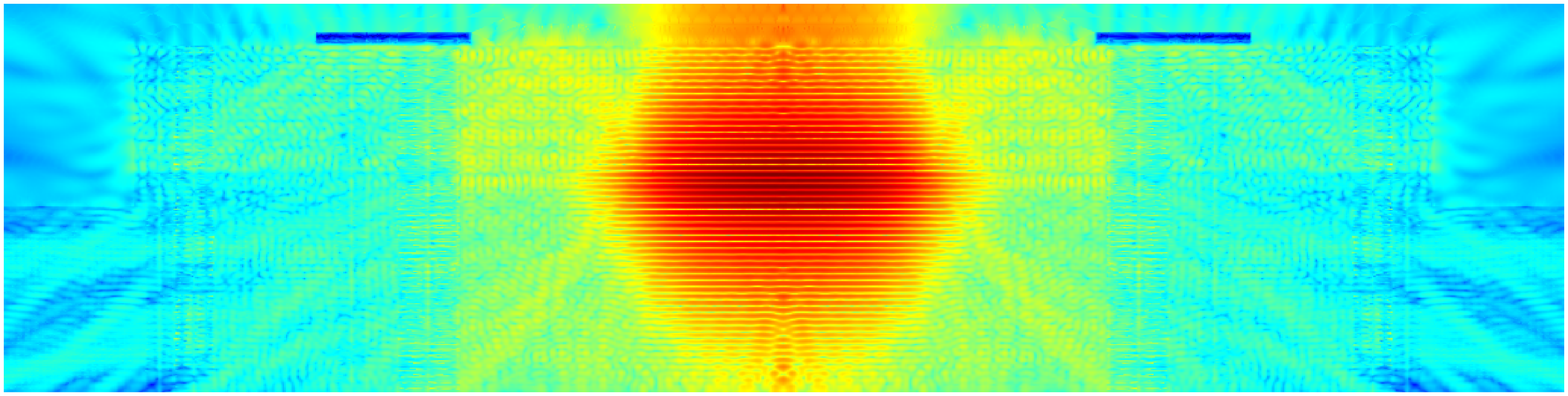}
\caption{\label{fig:vcselMode}Intensity of electric field obtained from FEM simulation in linear (top) and logarithmic (bottom) scale. In logarithmic scale the out-coupled field is visible.}
\end{figure}
The layout of the VCSEL example is shown in Fig. \ref{fig:vcselGeo}. The InGaAs active layer is embedded into a top and bottom GaAs/AlGaAs DBR mirror. Two AlOx appertures are placed above the active zone.  The corresponding material parameters were computed from \cite{AFR74} for AlGaAs and taken from Landolt-B\"ornstein for AlOx and InGaAs.

Since the design is cylinder symmetrical, simulations can be carried out on a 2D cross section. Figure \ref{fig:vcselMode} shows the fundamental mode of the VCSEL. On logarithmic scale we observe that the resonating mode radiates to the exterior. The corresponding resonance frequency is given by:
\begin{align*}
  \omega_{\mathrm{3D}}&=1.926\cdot 10^{15}+1.86\cdot 10^{11}i ,
\end{align*}
which corresponds to a wavelength of:
\begin{align*}
  \lambda_{\mathrm{3D}}&=978.12\,\mbox{nm}.
\end{align*}
The Bragg mirror was designed for a wavelength of $980\,$nm. 

\begin{figure}[]
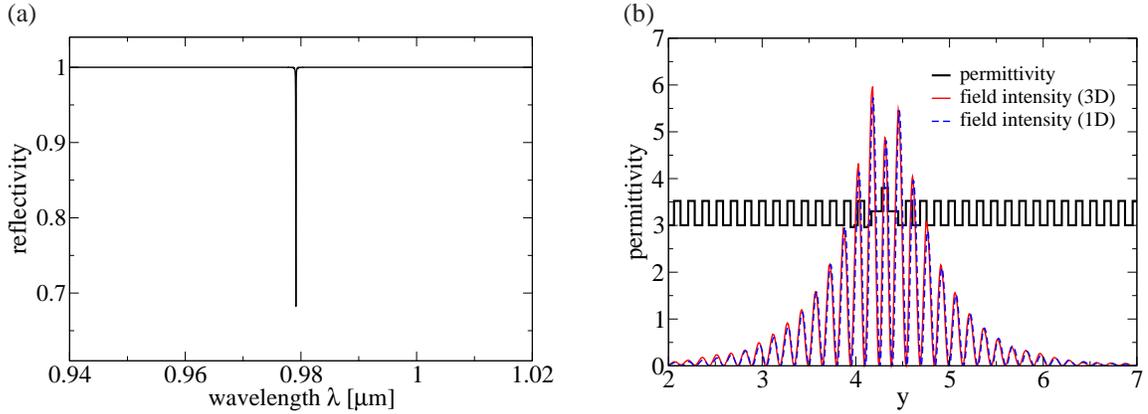

(a)\hspace{7.8cm}(b)\hfill\\\noindent
\includegraphics[height=5cm,clip]{refl1DVCSEL}\hfill
\includegraphics[height=5.1cm,clip]{fieldCut}
\caption{\label{fig:vcselFieldCut}(a) Reflectivity of 1D VCSEL model in dependence on incidence wavelength. The permittivity profile is given in (b): permittivity profile and electric field intensity of the fundamental mode along symmetry the axis of the VCSEL. A comparison of 1D semi-analytic and 3D finite element simulation is shown.}
\end{figure}

For the 1D computations a scattering problem for the dielectric profile along the VCSEL's symmetry axis was solved with varying wavelength of an incidence plane wave. The resonance wavelength was then determined from a reflectivity dip of the 1D profile, see Fig. \ref{fig:vcselFieldCut}(a). We found:
\begin{align*}
  \lambda_{\mathrm{1D}}&=979.12\,\mbox{nm},
\end{align*}
which is $1\,$nm apart from the 3D value. For comparison of the 3D model to 1D calculations, we extract the finite element solution along the symmetry axis of the computational domain. Figure \ref{fig:vcselFieldCut}(b) shows the corresponding permittivity profile along this axis together with the electric field intensity from 1D and 3D simulations. The near fields agree to a large extend. Of course the 1D simulations do not give the lateral mode shape and can not be used for lateral design of the VCSEL.
\subsection{Super large optical waveguide laser}
\label{sec:slow}
The SLOW laser fabricated  at the FBH was designed for a wavelength of $\lambda=1060\,$nm, which was used in numerical computations. It consists of a $9~ \mu$m broad vertical waveguide core in the middle of which the active region is placed and a wide ridge waveguide defined by lateral trenches filled with electroplated gold. The relative permittivities of the waveguide used in our simulations was $\epsilon=11.7$. The permittivities for the top and bottom claddings was fixed at $\epsilon=11.3$. More details on the design and experimental results can be found in \cite{aga}. The widths of both the vertical and the lateral waveguides are by a factor of nearly 10 larger than is commonly used. The design parameters are, e.g., the width w of the ridge and the residual layer thickness $d$, which denotes the distance between each trench and the active layer. Etching the trenches deeper into the material, reduces the residual layer thickness $d$.
\begin{figure}[]
\centering
\includegraphics[width=13cm]{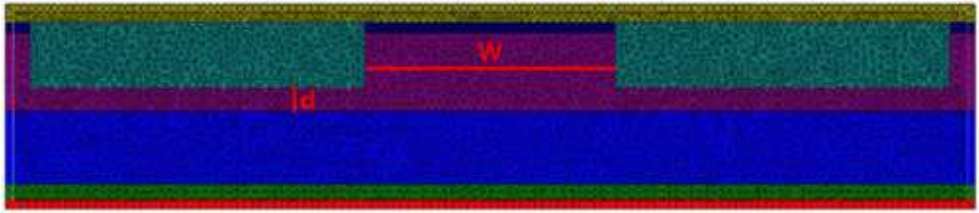}
\caption{\label{fig:geoSLOW}Finite element triangulation of cross section of super large optical waveguide laser. The meaning of the parameters w and $d$ is also indicated.}
\end{figure}
The finite element triangulation of our second example is shown in Fig. \ref{fig:geoSLOW}.

\begin{figure}[]
\centering
\includegraphics[width=8cm,clip]{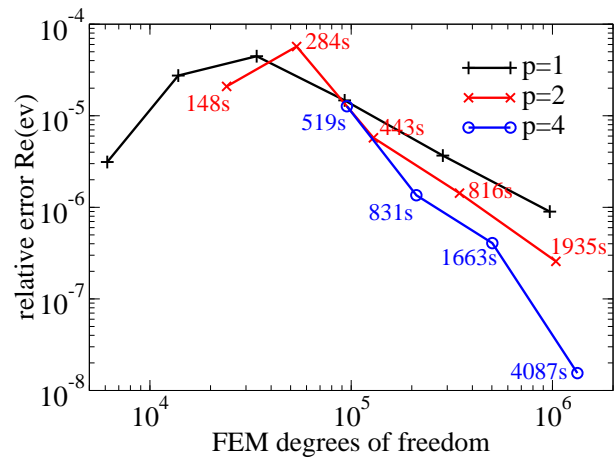}\hfill
\caption{\label{fig:conv}Convergence of fundamental eigenvalue for SLOW laser example. The relative error against a very accurate solution (computed with higher spatial discretization and $p=4$) is shown in dependence on number of finite element degrees of freedom (uniform refinement of triangulation) for different finite element orders $p$. Single CPU times are given for orders 2 and 4 (the computations were performed on Dual-Core AMD Opteron Processors with 2.8 GHz)}
\end{figure}
Let us look at the accuracy and computational costs for this example. Figure \ref{fig:conv} shows the convergence of the effective refractive index for different finite element orders $p$ in dependence on the spatial discretization. The convergence for all orders $p$ was computed against a reference value of:
\begin{align*}
  n_{\mathrm{eff}}=3.422595157 + 1.82097\cdot 10^{-4}i.
\end{align*}
This value was obtained from the most accurate finite element solution with order $p=4$ and a spatial discretization leading to $N=3956993$ finite element degrees of freedom. We observe that for high order $p$, less finite element degrees of freedom are necessary, to obtain an accurate solution. However, for all orders the relative error of the eigenvalue is very small, below $10^{-4}$ and down to $2\cdot 10^{-8}$ for finest discretization. For order $p=1$ first the relative error rises, and the convergent regime starts at $N\approx 30000$. Figure \ref{fig:conv} also gives single CPU times, where 10 modes closest to an initial guess were computed. Within a few minutes, very accurate finite element solutions can be computed for this real-world example.

\begin{figure}[]
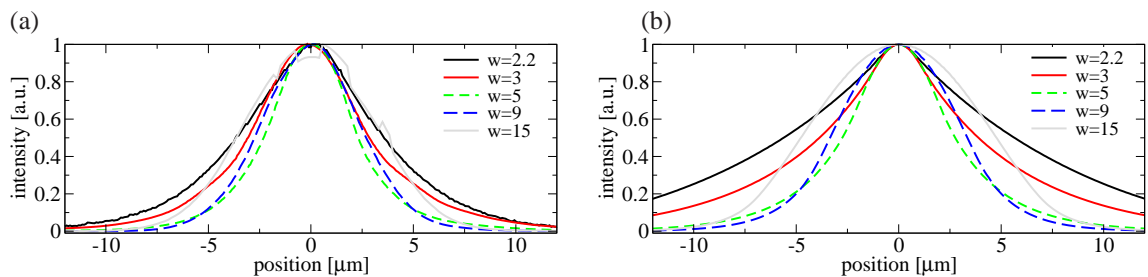

(a)\hspace{8cm}(b)\hfill\\\noindent
\includegraphics[width=7.3cm,clip]{exp}\hfill
\includegraphics[width=7.3cm,clip]{sim}
\caption{\label{fig:nearField}(a) Experimental and (b) numerical lateral near field "intensity profiles" of fundamental lasing mode for different values of w, given in $\mu$m.}
\end{figure}
\begin{figure}[]
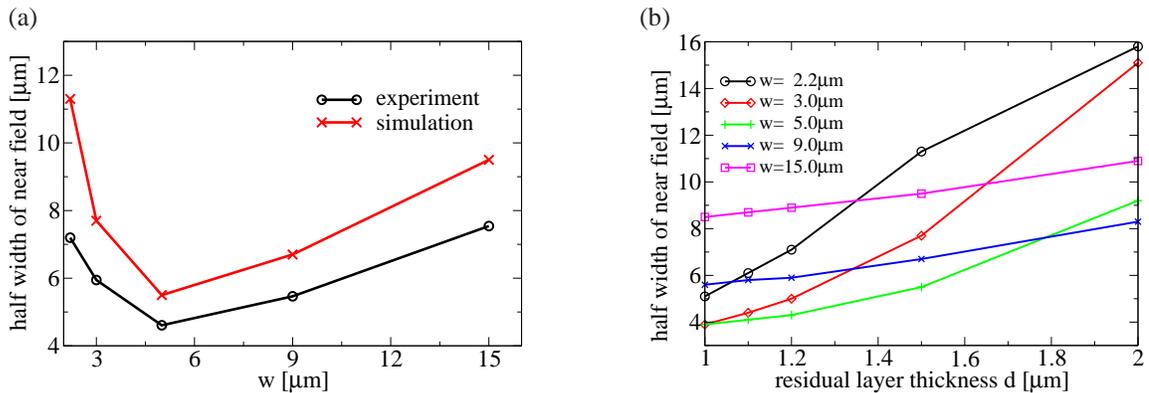

(a)\hspace{8cm}(b)\hfill\\\noindent
\includegraphics[height=4.7cm,clip]{comp}\hfill
\includegraphics[height=4.8cm,clip]{halfWidthsDRes}
\caption{\label{fig:scans}(a) Comparison of the widths (FWHM) of the experimental and numerical lateral near field intensity profiles corresponding to Fig. \ref{fig:nearField}; (b) widths of the numerical lateral near field intensity profiles in dependence on residual layer thickness $d$ for different values of w.}
\end{figure}
\begin{figure}[]
\centering
\includegraphics[height=2.8cm]{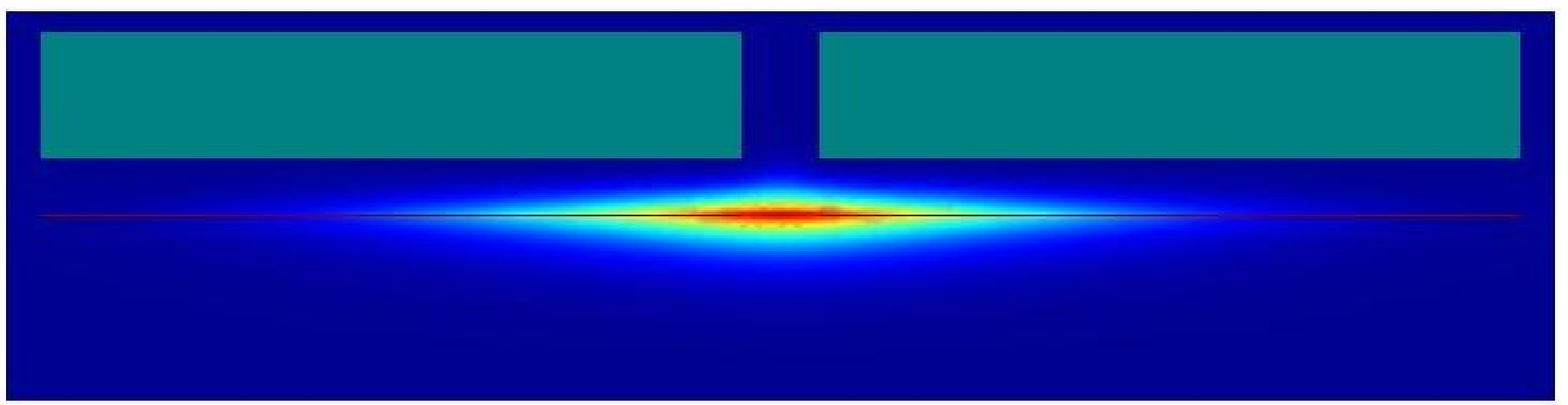}\vspace{0.5cm}
\includegraphics[height=2.8cm]{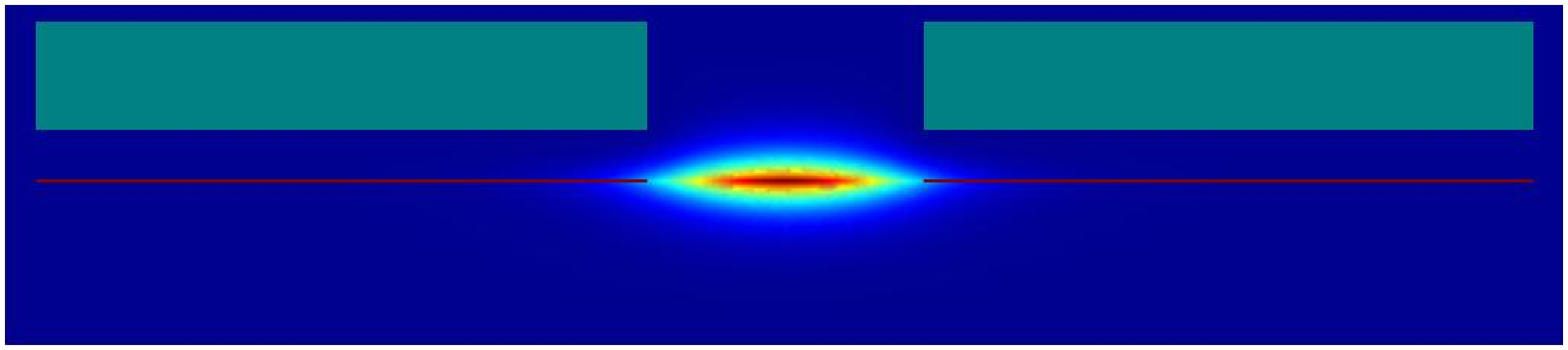}\vspace{0.5cm}
\includegraphics[height=2.8cm]{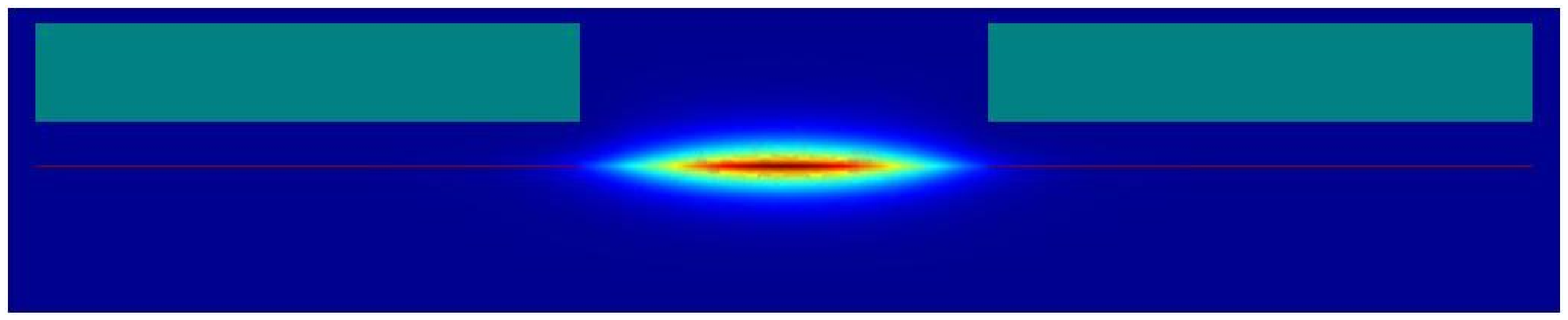}
\caption{\label{fig:nearFieldsFEM1}Intensity of fundamental lasing mode for varying $\mathrm{w}=2.2,\,9.0,\,15.0\mu$m. Gold trenches and active layer are shown in green and red respectively.}
\end{figure}

Figure \ref{fig:nearField}(a) shows corresponding experimental lateral near field profiles for different values of the design parameter w of the SLOW laser. The finite element results for the same design parameters are given in Fig. \ref{fig:nearField}(b), where the data are obtained from a lateral cut through the active zone. In order to compare experimental and numerical data, we extract the widths (FWHM, full width at half maximum) of the near field intensity profiles in dependence on the parameter w. The result is depicted in Fig.  \ref{fig:scans}(a). We observe a distinct minimum of the near field width for a value of $\mathrm{w} = 5~\mu$m. Although the experimental and numerical data show an offset, which is probably caused by the resolution limit of the measurement setup, the position of the minimum agrees. We observe that the lateral dimension of the lasing mode increases with increasing width of the ridge. However, reducing the width below $\mathrm{w} = 5~\mu$m leads to a broader near field distribution. In order to  explain this behaviour, the near field distributions obtained from finite element computation for different values of w are given in Fig. \ref{fig:nearFieldsFEM1}. We observe that for large w the lasing mode basically fills the ridge. For very small width the lasing mode, however, is no longer confined by the ridge and extends below the trenches towards the outer region. 

\begin{figure}[]
\centering
\includegraphics[height=2.8cm]{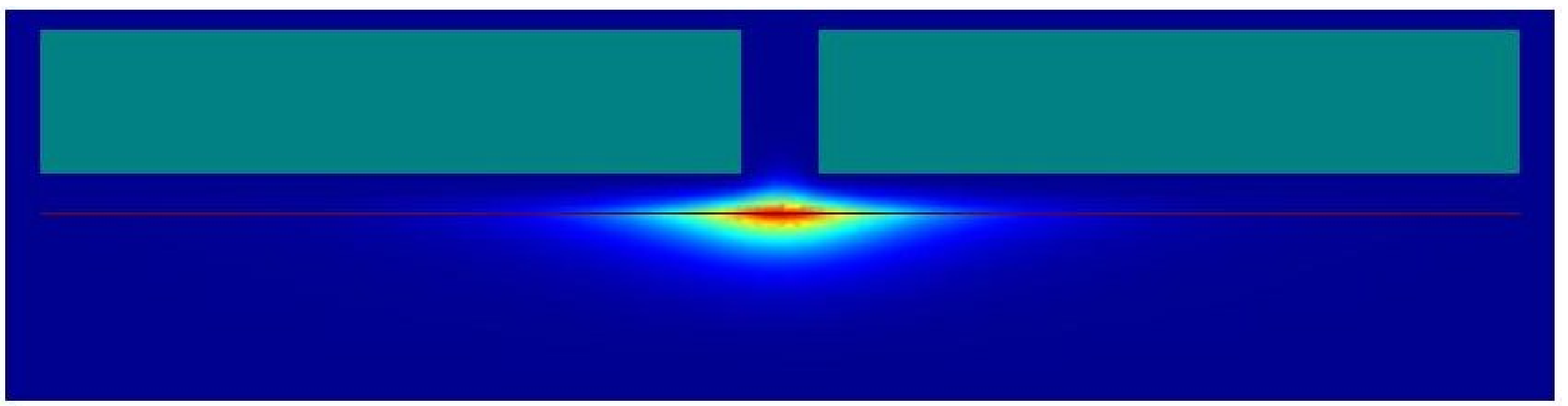}\vspace{0.5cm}
\includegraphics[height=2.8cm]{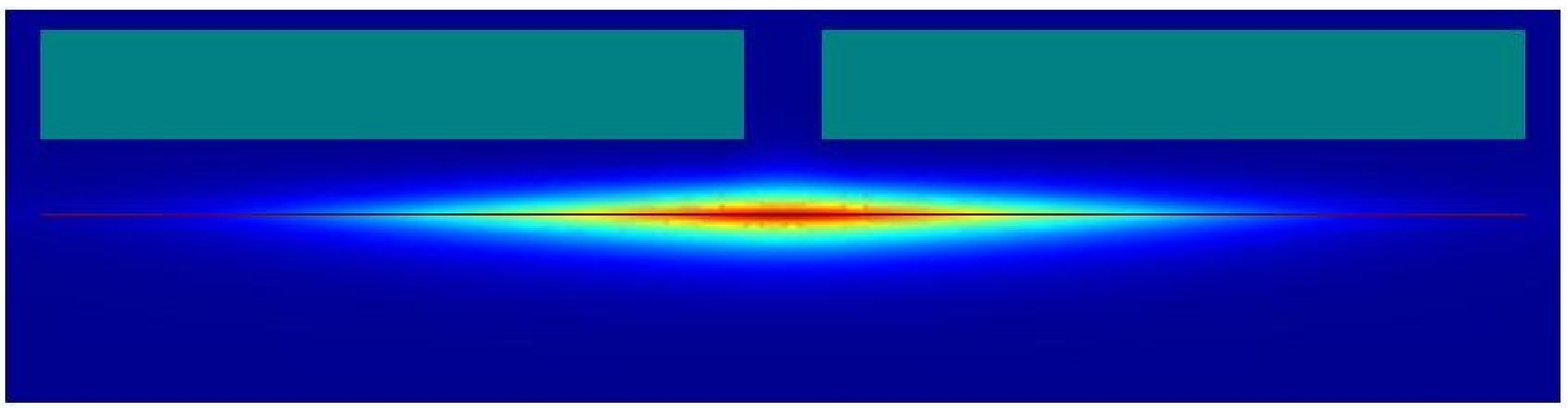}
\caption{\label{fig:nearFieldsFEM2}Intensity of fundamental lasing mode for fixed $\mathrm{w}=2.2\mu$m and varying residual layer thickness of $d=1.0,\,2.0\mu$m. Gold trenches and active layer are shown in green and red respectively.}
\end{figure}

In Figure \ref{fig:scans}(b) we analyze the dependence of the lateral width of the near field on the residual layer thickness $d$. We observe that for large w the residual layer thickness only has a weak influence on the width of the near field. However, for smaller values of w, a decreasing residual layer thickness leads to stronger lateral confinement of the lasing mode. Again finite element simulations help to understand this behaviour. Since for large w the lasing mode is mainly located between the trenches, c.f. Fig. \ref{fig:nearFieldsFEM1}, the residual layer thickness does not effect the mode. For a small value of w the mode extends below the trenches, and a decreasing residual layer thickness basically leaves no space for the mode below the trenches. The mode is then again confined between the trenches, see Fig. \ref{fig:nearFieldsFEM2}.

\section{Conclusions}
We showed that the finite element method is very well suited for the computation of optical modes in semiconductor lasers. A convergence analysis for an edge emitting laser demonstrated that accurate numerical results can be obtained at low computational costs. 

The numerical results were compared to real-world lasers. The dependence of experimentally obtained near field distributions on design parameters of the lasers could be reproduced and understood with the help of simulations.

Furthermore, the FEM results for a 3D VCSEL were compared to a simplified 1D model. The near field distributions of both models, thereby, agreed very well, whereas the computed eigenvalue differed slightly.

\section{Acknowledgment}
This work was carried out within SFB787 ``Halbleiter Nanophotonik'', funded by the DFG.

\bibliography{/home/numerik/bzfpompl/myBib}
\bibliographystyle{plain}

\end{document}